\documentstyle[prl,twocolumn,aps]{revtex}

  \input epsf
  \begin{document}
  \draft
  \title{Bloch oscillations and mean-field effects of Bose-Einstein
  condensates in 1-D optical lattices}
  \author{O. Morsch, J.H. M\"uller, M. Cristiani, D. Ciampini,
  and E. Arimondo}
  \address{INFM, Dipartimento di Fisica, Universit\`{a} di Pisa, Via
  Buonarroti 2, I-56127 Pisa, Italy}
  \date{\today}
  \maketitle
  \begin{abstract}
  We have loaded Bose-Einstein condensates into one-dimensional, off-resonant
  optical lattices and
  accelerated them by chirping the frequency difference between the two
  lattice beams. For small
  values of the lattice well-depth, Bloch oscillations were observed.
  Reducing the potential depth further, Landau-Zener tunneling out of
  the lowest lattice band, leading to a breakdown of the oscillations, was
  also studied and used as a probe for the effective potential resulting
from mean-field interactions
  as predicted by Choi and Niu [Phys. Rev. Lett. {\bf 82}, 2022 (1999)].
  The effective potential was measured for various condensate
  densities and trap geometries, yielding good qualitative agreement with
  theoretical calculations.
  \end{abstract}

  \narrowtext

  The properties of ultra-cold atoms in periodic light-shift potentials in
  one, two and
  three dimensions have been investigated extensively in the past ten
  years~\cite{lattreview}. In
  near-resonant and, more recently, far-detuned optical lattices, a variety
  of phenomena have been
  studied, such as the magnetic properties of atoms in optical lattices,
  revivals of wave-packet
  oscillations, and Bloch oscillations in accelerated
  lattices~\cite{raizen97}.
  While in most of the original optical lattice experiments the atomic clouds
  had temperatures in the
  the micro-Kelvin range, corresponding to a few recoil energies of the
  atoms, samples with
  sub-recoil energies are now routinely produced in Bose-Einstein
  condensation experiments. Many aspects of Bose-Einstein
  condensed atomic clouds
  (BECs) have been studied~\cite{inguscio}, ranging from collective
  excitations to superfluid
  properties and quantized vortices. So far, the majority of these
  experiments have been carried out
  essentially in harmonic-oscillator potentials provided by magnetic traps or
  optical dipole traps.
  The properties of BECs in periodic potentials constitute a vast new field of
  research (see, for
instance,~\cite{sorensen98,jaksch98,choi99,javanainen99,brunello00,zobay00,chiofalo00,trombettoni01,poetting01}).
  Several experiments in the pulsed standing wave
  regime~\cite{kozuma99,stenger99} as well as studies of the
  tunneling of BECs out of the potential wells of a shallow optical lattice
  in the presence of
  gravity~\cite{anderson98}, the creation of squeezed states in
condensates~\cite{orzel}, and the search
for superfluid
  dynamics~\cite{burger01} have taken the first steps in that direction. In
  this paper, we present
  the results of experiments on BECs of $^{87}\mathrm{Rb}$ atoms in
  accelerated optical lattices. In
  particular, we demonstrate coherent acceleration and Bloch oscillations
of BECs adiabatically
  loaded into optical
  lattices and the reduction of the effective potential seen by
  the condensates due to mean-field interactions. The latter was
  inferred from measurements of Landau-Zener tunneling when the lattice
depth was further reduced and/or
  the acceleration
  increased. We loaded the condensate into optical lattices with different
  spatial
  periods, generating the periodic optical lattice either from two
  counter-propagating laser beams
   or two laser beams enclosing an angle $\theta$ different from
$180\,\mathrm{deg}$.

   The properties of a Bose-Einstein condensate located in a periodic
optical
  lattice with depth $U_0$ are described through the Gross-Pitaevskii
  equation valid for the
  single-particle wavefunction~\cite{footnote_gp}.
 In agreement
  with the Bloch approach, the condensate excitation spectrum exhibits a
  band structure, and in the presence of an
  acceleration of the optical lattice Bloch oscillations of the condensate
  should occur~\cite{sorensen98,choi99}. We present experimental evidence for
Bloch
  oscillations preserving the condensate wavefunction. The nonlinear
interaction of the condensate
  may be described through a dimensionless
  parameter~\cite{choi99,chiofalo00,trombettoni01} $C=g/E_{\rm B}$
corresponding to the ratio
  of the nonlinear interaction
  term $g=4\pi n_0 \hbar^2 a_{s}/M$ and the lattice Bloch energy
  $E_{\rm B}=\hbar^2(2\pi)^2/Md^2$. The parameter $C$ contains the
  peak condensate density $n_0$, the s-wave scattering
  length $a_{s}$, the atomic mass $M$, the lattice constant $d=\pi
  /\sin(\theta/2)k$,
  with $k$ the laser wavenumber, and $\theta$ the angle between the two
  laser beams creating the
  1-D optical lattice. From this it follows that a small angle $\theta$
should result in a large
  interaction term $C$. In fact, creating a lattice with
  $\theta = 29\,\mathrm{deg}$ allowed us to
  realize a value of $C$ larger by a factor of more than 10 with respect
  to~\cite{anderson98} using a
  comparable condensate density. In the following, the parameters $d$,
  $E_{\rm B}$ and $C$ always refer to the respective lattice
  geometries with angle $\theta$.

  The role of the
  nonlinear interaction
  term of the Gross-Pitaevskii equation may be
  described through an effective potential in a non-interacting gas
model~\cite{choi99,chiofalo00}.
  In the perturbative regime of~\cite{choi99} the effective potential
  is
  \begin{equation}
  U_{\rm eff}=U_0/(1+4C),
  \label{pot}
  \end{equation}
  so that the potential seen by the condensate is $U(x)= U_{\rm
eff}\sin(2\pi x/d)+\mathrm{const.}$ We therefore expect that for
large values of $C$,
  i.e. large mean-field
effects, the effective optical lattice potential acting on the
condensate should be
  significantly
  reduced.
\begin{figure}
  \centering\begin{center}\mbox{\epsfxsize 2.8 in \epsfbox{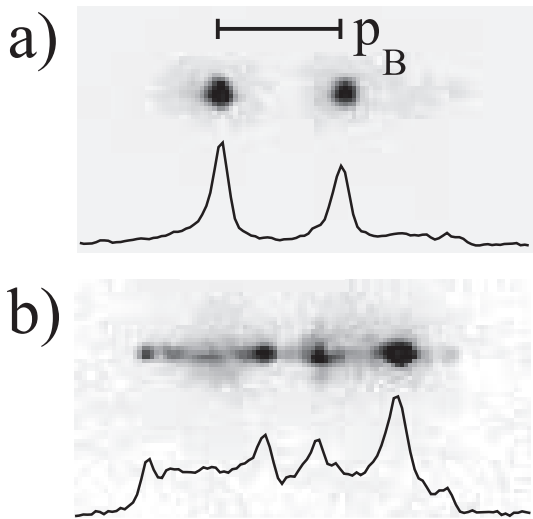}}
  \end{center}
  \caption{Condensate acceleration in the counter-propagating optical
  potential. Shown here are typical time-of-flight images for moderate,
  $n_0<10^{14}\,\mathrm{cm^{-3}}$, in (a)
   and high condensate density, $n_0> 2 \times
   10^{14}\,\mathrm{cm^{-3}}$, in
   (b). The two-peaked
  coherent structure in (a) is easily interpretable within the
  periodic potential diffraction picture giving rise to two momentum classes,
  whereas the more complicated pattern in (b) cannot be described by this
simple model.}
  \label{profiles}
  \end{figure}
  Our apparatus used to achieve Bose-Einstein condensation of
  $^{87}\mathrm{Rb}$ is described in detail in~\cite{jphysbpaper}.
  Essentially, $~5\times
  10^7$ atoms captured in a magneto-optical trap (MOT) were transferred into
  a triaxial
  time-orbiting potential trap (TOP)~\cite{prlpaper}. Subsequently, the atoms
  were evaporatively
  cooled down to the transition temperature for Bose-Einstein condensation,
  and after further cooling
  we obtained condensates of $\approx 10^4$ atoms without a discernible thermal
  component in a magnetic trap with frequencies around
$15-30\,\mathrm{Hz}$. In one
  set of experiments, the magnetic trap was then switched off and a
  horizontal 1-D optical lattice was switched on, while in the other case the
  interaction between the
  condensate and the lattice took place inside the magnetic trap, which was
  subsequently switched off
  to allow time-of-flight imaging. The lattice direction was parallel to the
strong
   axis of the trap, and the lattice beams were created by a
  $50\,\mathrm{mW}$ diode
  slave-laser injected by a grating-stabilized master-laser blue-detuned by
  $\Delta\approx
  28-35\,\mathrm{GHz}$ from the $^{87}\mathrm{Rb}$ resonance line. After
  passage through an optical
  fibre, the laser light was split and passed through two acousto-optic
  modulators (AOMs) that were
  separately controlled by two phase-locked RF function generators operating
  at frequencies around
  $80\,\mathrm{MHz}$, with a frequency difference $\delta$. The first-order
output beams of the
  AOMs generated the
  optical lattice, and an acceleration of the lattice was effected by applying
  a linear ramp to $\delta$.

  For the values of the detuning and
  laser
  intensity used in our
  experiment, the spontaneous photon scattering rate ($\approx
  10\,\mathrm{s^{-1}}$) was negligible
  during the interaction times of a few milliseconds. In our experimental
  setup, we
  realized a counter-propagating lattice geometry with
  $\theta=180\,\mathrm{deg}$
   and an angle geometry with $\theta=29\,\mathrm{deg}$, leading to
   lattice constants $d$ of  $0.39$ and $1.56\,\mathrm{\mu m}$,
   respectively~\cite{footnote}. Inserting a peak
  condensate density of $n_0 \approx
  10^{14}\,\mathrm{cm^{-3}}$ (typical of trap frequencies $>
100\,\mathrm{Hz}$) in the
Thomas-Fermi limit into the expression for the interaction
parameter $C$ leads to $C=0.06$
  for the counter-propagating configuration and $C=0.25$ for
$\theta=29\,\mathrm{deg}$.
    In order to determine the lattice depth at low condensate
   densities, we measured the Rabi frequency on a first order Bragg
  resonance~\cite{kozuma99}. Typically, for well-aligned lattice
  beams we measured lattice depths up to $20\%$ lower than the
  theoretically expected value inferred from the laser intensity and
  detuning. These discrepancies are within the calibration error of
  our laser power measurements.

In order to accelerate the
condensate,
   we adiabatically loaded it into the
  lattice by switching one of the lattice beams on suddenly and ramping the
  intensity of the other
  beam from $0$ to its final value in $~200\,\mathrm{\mu s}
  $~\cite{footnote1}. Thereafter, the
  linear increase of the detuning $\delta$ provided a constant acceleration
  $a=\frac{\lambda}{2\sin(\theta/2)}\frac{d\delta}{dt}$ of the optical
lattice, leading to a final lattice velocity
  $v_{\rm lat}=\frac{\lambda}{2\sin(\theta/2)}\delta_{f}$, where
$\delta_{f}$ is the
final detuning between the beams. After a few
  milliseconds of acceleration, the lattice beams were switched off and the
  condensate was imaged
  after another $10-15\,\mathrm{ms}$ of free fall. As the lattice can
  only transfer momentum to the atoms in the
  condensate in units of the Bloch momentum $p_{\rm B}=\hbar (2\pi/d)$, the
  acceleration of the condensate showed up as diffraction
  peaks corresponding to higher momentum classes as time increased. Since
for our
   magnetic trap parameters the initial momentum spread of the condensate
(which is transferred into a spread of the lattice quasimomentum
   during an adiabatic switch-on) was
  much less than a
  recoil momentum of the optical lattice, the different
  momentum classes
  $|p=\pm np_{\rm B}\rangle$ (where $n=0,1,2,...$) occupied by the
  condensate wavefunction could be resolved directly after the
  time-of-flight (see, for instance, the peaks corresponding to $n=0$ and
$n=1$ in
  Fig.~\ref{profiles}(a)).  As described in~\cite{bendahan96}, the acceleration
  process within a periodic potential can also be
  viewed as a succession of adiabatic rapid passages between momentum states
  $|\pm np_{\rm B} \rangle$. We observed a momentum transfer
  of up to $6\,p_B$ without a detectable reduction of the
  phase-space density of the condensate.
  We verified that in the process of the
acceleration and tunneling,
   the condensed fraction was not reduced for low condensate densities. Our
investigation did not, however, test the evolution
  of the condensate phase, but on the basis of the Bragg scattering
experiments of~\cite{kozuma99}
  we assume that the interaction times of our
  experiment should not destroy the condensate phase.
 \begin{figure}\centering\begin{center}\mbox{\epsfxsize 3.1 in \epsfbox{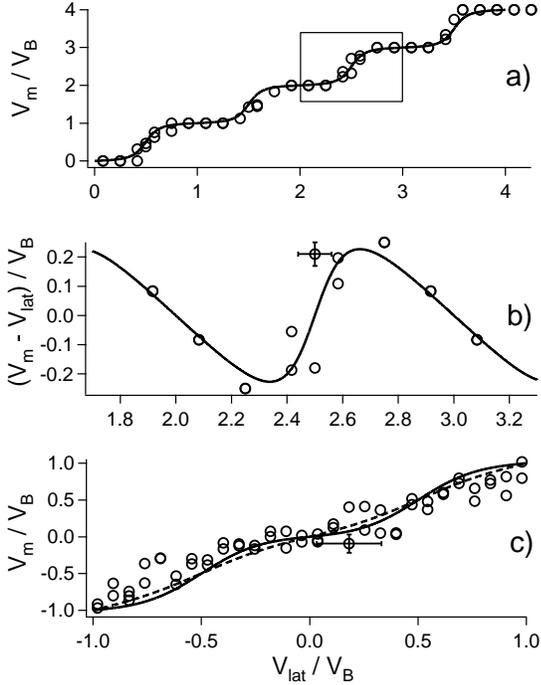}}
  \end{center}
  \caption{Bloch oscillations of the condensate mean
  velocity $v_{m}$ in an optical
  lattice. (a) Acceleration
  in the counter-propagating lattice with $d=0.39\,\mathrm{\mu m}$, $U_0\approx
  0.29\,E_{\rm B}$ and
  $a=9.81\,\mathrm{m\,s^{-2}}$. Solid line: theory. (b) Bloch oscillations
  in the rest frame of the
  lattice, along with the theoretical prediction (solid line) derived from
  the shape of the lowest
  Bloch band. (c) Acceleration in a lattice with $d=1.56\,\mathrm{\mu
  m}$,
  $U_0\approx
  1.38\,E_{\rm B}$ and $a=0.94\,\mathrm{m\,s^{-2}}$. In this case, the
Bloch oscillations are much less pronounced.
  Dashed and solid lines:
  theory for $U_0=1.38\,E_{\rm B}$ and $U_{\rm eff}\approx 0.88\,E_{\rm B}$.}
  \label{fig3}
  \end{figure}
  The average velocity of
  the condensate was derived from the occupation numbers of the different
momentum
  states. Figure~\ref{fig3}
  shows the results of the acceleration of a
  condensate  in the counter-propagating lattice with $U_{\rm
eff}=0.29\,E_{\rm B}$ and
  $a=9.81\,\mathrm{m\,s^{-2}}$. In the rest-frame of the
  lattice (Fig.~\ref{fig3} (b)), one clearly sees Bloch oscillations of the
  condensate velocity
  corresponding to a Bloch-period
  $\tau_{B}=\frac{h}{M_{Rb}ad}=1.2\,\mathrm{ms}$. The shape of
  these oscillations agrees well with the theoretical curve calculated from
  the lowest energy band of
  the lattice.

 When the experiment was repeated in the angle-configuration
  of the lattice, larger values of the lattice depth in units of
  the Bloch energy $E_B$ could be realized. Because
  of the reduced Bloch
  velocity $v_{\rm B}$ in this geometry, the acceleration process was
  extremely sensitive
  to any initial velocity of the condensate, which in our TOP trap is
  intrinsically given by the
  micromotion~\cite{prlpaper} at the frequency of the bias field. For the
  trap parameters used in our
  experiments, the velocity amplitude of the micromotion could be of the same
  order of magnitude as
  $v_{\rm B}$ and the condensates could, therefore, have quasimomenta close
  to the edge of the
  Brillouin zone. In order to counteract this, we
  performed the acceleration
  experiments
  inside the magnetic trap, eliminating the velocity of the condensate
  relative to the lattice by
  phase-modulating one of the lattice beams at the same frequency and in
  phase with the rotating bias
  field of the TOP trap. In this way, in the rest frame of the lattice the
  micromotion was
  compensated. Nevertheless, a residual sloshing of the condensate with
  amplitudes $<3\,\mathrm{\mu m}$
  could not be ruled out, so that the uncertainty in the initial velocity of
  the condensate was still
  around $0.5\,\mathrm{mm\,s^{-1}}$, corresponding to $\approx
  0.15\,v_{\rm B}$ in this geometry. Fig.~\ref{fig3} (c)
  shows the results of a measurement of the acceleration of the
  condensate as a function of the final lattice velocity for a
  theoretical lattice depth of $U_0\approx 1.38\,E_{\rm B}$ together with the
theoretical
  curves for the same potential
   and the (assumed) effective potential $U_{\rm
  eff}\approx0.65\,U_0\approx
  0.88\,E_{\rm B}$ as calculated from the condensate density
  using the perturbative expression of Eq.~(\ref{pot}).
  As expected from the band-structure calculations, the Bloch
  oscillations are much less pronounced in this geometry.
 \begin{figure}\centering\begin{center}\mbox{\epsfxsize 3.1 in \epsfbox{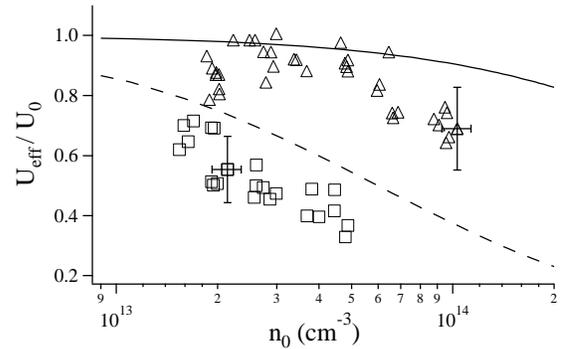}}
  \end{center}
  \caption{Dependence of the effective potential $U_{\rm eff}$ on the
  peak
   density $n_0$ for the two lattice geometries. The experimental
results for the counter-propagating
   (triangles) and angle geometries (squares) are plotted together with the
theoretical
   predictions
(solid and dashed lines, respectively). Parameters in these
experiments were $a=23.4\,\mathrm{m\,s^{-2}}$ and
$U_0=0.28\,E_{\mathrm{B}}$ for the counter-propagating lattice and
$a=3.23\,\mathrm{m\,s^{-2}}$ and $U_0=0.71\,E_{\mathrm{B}}$ for
the angle geometry.}
  \label{meanfield}
  \end{figure}
  In order to measure more accurately the variation of the
  effective potential $U_{\rm eff}$ with the interaction parameter
  $C$, we studied Landau-Zener tunneling out of the lowest Bloch
  band for small lattice depths in both geometries. To this end,
  the acceleration of the lattice was increased in such a way that
  when the condensate crossed the edge of the Brillouin zone, an
  appreciable fraction of the atoms tunneled across the band gap
  into the first excited band (and,
  therefore, effectively
  to the continuum, as the gaps between higher bands are negligible for the
  shallow potentials used
  here). According to Landau-Zener theory, this fraction is~\cite{bendahan96}
  \begin{equation}
   r=\exp\left(-{\frac{\pi U_{\rm eff}^2}{8\hbar p_{B}a}}\right),
   \end{equation}
   giving a velocity $v_m=(1-r)v_{\rm B}$
of the condensate
  at the end of the acceleration process for a final velocity $v_{\rm B}$ of
  the lattice~\cite{footnote3}. We verified that this formula
  correctly described the tunneling of the condensate in our
  experiment by varying both the potential depth and
  acceleration.

  Thereafter, we studied the variation of the final mean velocity
  $v_m$ as a function of the condensate density for the two
  lattice geometries. The density was varied by changing the mean frequency
of the magnetic trap (from
  $\approx 25\,\mathrm{Hz}$ to $\approx 100\,\mathrm{Hz}$). From
  the mean velocity the effective potential was then calculated using the
Landau-Zener
  formula given above. Fig.~\ref{meanfield} shows the ratio $U_{\rm eff}/U_0$
  as a function of the peak density $n_0$ for the
  counter-propagating geometry and the angle-geometry.
  As expected, the reduction of the effective potential is much
  larger in the angle geometry. The theoretical predictions of
  Eq.~(\ref{pot}) are also shown in the figure, with the
  potential $U_{0}$ calculated taking into account
  losses at the cell windows and imperfections of the
  polarizations of the lattice beams. The residual combined error
  due to uncertainties in the absolute intensity measurements,
  the position of the beam axes relative to the position of the
  BEC and a small initial velocity due to sloshing of the BEC in the
magnetic trap, as well as a systematic error due to the difference
in
  position of about $300\,\mathrm{\mu m}$ of the condensate
  between the weakest and the strongest trap used, was estimated
  at about $20 \%$. Within these experimental uncertainties,
  qualitative agreement with theory was good~\cite{footnote_n}. Although we
could
  realize peak
  densities up to $4\times 10^{14}\,\mathrm{cm^{-3}}$ by using larger trap
frequencies, data
  points for $n_0> 10^{14}\,\mathrm{cm^{-3}}$ in the
counter-propagating lattice
  and $n_0>5\times 10^{13}\,\mathrm{cm^{-3}}$ in the angle geometry were
not included
  in the graph as the resulting diffraction patterns were not
  easily interpretable within the simple model described above (see
  Fig.~\ref{profiles}(b)).

  In summary, we have investigated the coherent acceleration and
Bloch oscillations of
  Bose-Einstein condensates adiabatically loaded into a 1-D optical
  lattice. Through Landau-Zener tunneling out of the lowest Bloch
  band, we have studied the dependence of the effective potential
  on the interaction parameter $C$. The
  results obtained are in
  good qualitative agreement with the available theories and extend the
experimental
work on ultra-cold atoms in optical
  lattices into the domain of Bose-Einstein condensates. In order to
improve the theoretical
  description of our experiment, the finite extent of the condensate leading to
  the occupation of only a few lattice sites and the three-dimensional nature
   of the condensate evolution as well as the role of the interaction term
in the adiabaticity criterion for switching on the lattice
   will have to be
  taken into account.

  This work was
  supported by the MURST through the PRIN2000 Initiative, by the INFM through the Progetto di Ricerca Avanzata `Photon Matter', and by the
  EU through the Cold Quantum-Gases Network, contract
  HPRN-CT-2000-00125. O.M. gratefully acknowledges a
  Marie-Curie Fellowship from the EU within the IHP Programme. The
  authors thank M. Anderlini for help in the data acquisition.


\begin{references}
\bibitem{footnote_gp}A single mode description leading to the
Gross-Pitaveskii equation is justified for our lattice depths well
below the tight binding limit.
\bibitem{footnote} A large lattice constant also means that only a
  few potential wells are occupied by the condensate.
\bibitem{footnote1} The adiabaticity criterion of ref.~\cite{bendahan96}
  refers to interband transitions
  due to the rate of change of the band energies of non-interacting
  particles. This criterion was
  extended in~\cite{choi99} to the condensate on the basis of the
  band energies for the
  effective potential. However, other criteria such as collective
  excitations
  of the condensate~\cite{sorensen98} or the perturbation of the global
condensate phase (\cite{orzel} and C. Williams, private
communication), may impose more stringent conditions but have not
  been considered here. In the Landau-Zener tunneling experiments,
  we extended the ramping time to $3\,\mathrm{ms}$.
\bibitem{footnote3}These expressions are only exact if the
  final velocity of the lattice is an integer multiple of $v_{\rm
  B}$, as in that case the mean velocity of the condensate is
  independent of the band structure.
\bibitem{footnote_n} The theoretical analysis of
ref.~\cite{choi99} assumes a uniform condensate density, while in
our experiment there is a spatial dependence of the density due to
the harmonic trapping potential. In calculating the theoretical
values for $U_{\rm eff}/U_0$ through the parameter $C$, we used the
peak density of the condensate.
\end{references}
  \end{document}